\documentclass[twocolumn,showpacs,preprintnumbers,amsmath,amssymb]{revtex4}

\usepackage{graphicx}

\date{\today}

\begin{document}

\title{Entanglement induced by nonadiabatic chaos}

\author{Hiroshi Fujisaki}\email{fujisaki@bu.edu}
\affiliation{
Department of chemistry,
Boston University, 590 Commonwealth Ave.,
Boston, 02215, Massachusetts, USA
}

\begin{abstract}

We investigate entanglement
between electronic and nuclear degrees of freedom 
for a model nonadiabatic system.
We find that entanglement 
(measured by the von Neumann entropy of the subsystem for the eigenstates) 
is large in a statistical sense 
when the system shows ``nonadiabatic chaos'' behavior 
which was found in our previous work 
[Phys.~Rev.~E {\bf 63}, 066221 (2001)]. 
We also discuss non-statistical behavior of the eigenstates 
for the regular cases.

\end{abstract}


\maketitle

Quantum information processing (QIP)
is one of hot topics in many branches of science \cite{NC00}.
One important point is how to implement a quantum 
computer in real systems, and many possibilities 
have been theoretically suggested and experimentally tested.
One candidate can be molecular systems because 
highly excited molecules have dense quantum states, 
which can be manipulated by laser fields.
Some quantum logic gates in such a system 
can be built by using optimal control theory \cite{PK02},
and are actually realized in a molecular system \cite{VAZLK02}.

In highly excited molecules or laser-driven molecular systems,
nonadiabatic transition (NT)
is a rule rather than an exception \cite{Nakamura02,TAWM00},
i.e., 
we have to consider not only electronic or nuclear degrees of freedom (DoF), 
but both at the same time.
In such a case, a fundamental issue related to QIP
is how much (quantum) entanglement is produced between 
electronic and nuclear DoF in molecular systems,
because entanglement is a key ingredient in QIP.

Here we investigate 
a two-mode-two-state (TMTS) system which has two electronic DoF and two 
nuclear (vibrational) DoF with a nonadiabatic coupling \cite{FT01b,Heller90}. 
This is a (minimum) model NT system which shows ``quantum chaos'' behavior,
i.e., statistical properties of energy levels and eigenstates
are similar to those of a random matrix system \cite{Gutzwiller90}. 
If many electronic DoF are involved, 
the similar system has a naive 
classical limit, and its dynamical property of entanglement has 
been already addressed in \cite{FNP98}.
In such a case, a quantum chaological view is effective,
and we can say much about a quantum system by studying its 
classical limit \cite{FNP98}.
However, the situation is different and more difficult here,
because the TMTS system does not have a naive classical limit 
due to discreteness of the electronic DoF \cite{Fujisaki03}, 
and deserves further attentions in view of entanglement.

\begin{figure}[htbp]
\hfill
\begin{center}
\includegraphics[scale=0.4]{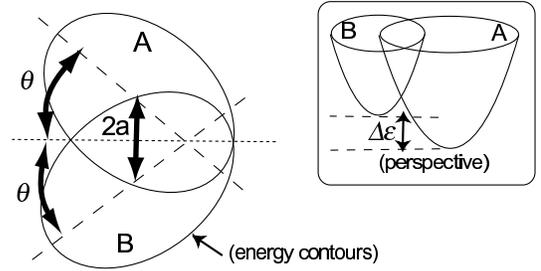}
\caption{
\label{fig:TMTS}
A shematic representation of the TMTS system. The distance between the minima of the 
potential is $2 a$, and the angle between the relevant crossing seam (dotted line) 
and the primary axis of each potential (dashed line) is $\theta$.
Inset: The perspective of the TMTS system. The potential minima are 
different with $\Delta \epsilon=\epsilon_B-\epsilon_A=0.173$.
}
\end{center}
\end{figure}

The TMTS system in the diabatic representation 
is described by the following Hamiltonian:
\begin{equation}
{\cal H}_{\rm TMTS}=\left( 
\begin{array}{cc}
T_{{\rm kin}}+V_{A} & J \\ 
J & T_{{\rm kin}}+V_{B}
\end{array}
\right),  
\label{chaos_Heller}
\end{equation}
where $T_{\rm kin}$ is the kinetic energy,
$V_i$ ($i=A,B$) is the potential energy for state $i$
defined by
\begin{eqnarray}
T_{{\rm kin}} 
&=& \frac{1}{2}(p_{x}^{2}+p_{y}^{2}), 
\\
V_i 
&=& 
\frac{1}{2}(\omega _{x}^{2} \xi_i^{2}+\omega _{y}^{2} \eta_i^{2})+\epsilon_i 
\quad (i=A,B)
\end{eqnarray}
with
\begin{eqnarray}
\xi_A 
&=& 
x \cos \theta -(y- a) \sin \theta,  
\\
\eta_A 
&=& 
x \sin \theta +(y- a) \cos \theta,  
\\
\xi_B
&=& 
x \cos \theta +(y+ a) \sin \theta,  
\\
\eta_B 
&=& 
-x \sin \theta +(y+ a) \cos \theta.  
\label{eq:xieta}
\end{eqnarray}
The geometrical meaning of the parameters is shown 
in Fig.~\ref{fig:TMTS}.
Note that we have just used a harmonic potential for each state.

Here the Duschinsky angle $\theta$ \cite{TLL03} 
and the nonadiabatic coupling constant $J$ are two 
important parameters for the system; The latter induces entanglement 
between electronic and vibrational DoF.
We solve this Hamiltonian according to the procedure in \cite{FT01b},
and obtain the eigen-energies and eigen-vectors.
The $k$-th eigenvector can be written as 
\begin{equation}
|\Phi^{(k)} \rangle
=\sum_{i=1}^2 \sum_{j} C^{(k)}_{ij} |i \rangle \rangle | j \rangle
=\sum_{i=1}^2 |\phi_i^{(k)} \rangle |i \rangle \rangle
\end{equation}
where $|1 \rangle \rangle = (|A \rangle +|B \rangle)/\sqrt{2}$,
$|2 \rangle \rangle = (-|A \rangle +|B \rangle)/\sqrt{2}$,
$|A \rangle$, $|B \rangle$ are the electronic 
bases for diabatic surfaces $A$ and $B$, respectively,
$|j \rangle$ represents 2D harmonic eigenfunctions,
and $|\phi_i^{(k)} \rangle \equiv \langle \langle i |\Phi^{(k)} \rangle
=\sum_{j} C^{(k)}_{ij} | j \rangle$.

From this eigenvector, we can construct a reduced density 
operator for the electronic DoF as 
\begin{eqnarray}
\rho^{(k)}
&=& {\rm Tr}_{\rm vib} \{ |\Phi^{(k)} \rangle \langle \Phi^{(k)}| \}
=\left(
\begin{array}{cc} 
\rho^{(k)}_{11} &  \rho^{(k)}_{12}
\\ 
\rho^{(k)}_{21} &  \rho^{(k)}_{22}
\end{array}
\right)
\nonumber
\\
&=& 
\left(
\begin{array}{cc} 
\sum_j C_{1,j}^{(k)} (C_{1,j}^{(k)})^*
 &  \sum_j C_{1,j}^{(k)} (C_{2,j}^{(k)})^*
\\ 
\sum_j C_{2,j}^{(k)} (C_{1,j}^{(k)})^*
 &  \sum_j C_{2,j}^{(k)} (C_{2,j}^{(k)})^*
\end{array}
\right)  
\label{eq:ele_den}
\end{eqnarray}
where $C_{i,j}^{(k)}$ are actually all real numbers.

The measure of entanglement we choose here is the von Neumann 
entropy of the subsystem defined by
\begin{eqnarray}
S^{(k)}_{\rm vN} 
&=& -{\rm Tr} \{ \rho^{(k)} \log \rho^{(k)} \}
\nonumber
\\
&=&
-\lambda_1^{(k)} \log \lambda_1^{(k)}
-\lambda_2^{(k)} \log \lambda_2^{(k)} 
\end{eqnarray}
where $\lambda_i^{(k)}$ ($i=1,2$) is an eigenvalue for the 
2$\times$2 matrix, Eq.~(\ref{eq:ele_den}).
A note is in order: 
the value of the entropy is the same if we use the reduced density 
operator for the vibrational DoF. We took the electronic DoF because 
the 2 $\times$ 2 matrix is very easy to diagonalize,
and to interpret the result as shown below.

\begin{figure}[htbp]
\hfill
\begin{center}
\includegraphics[scale=0.6]{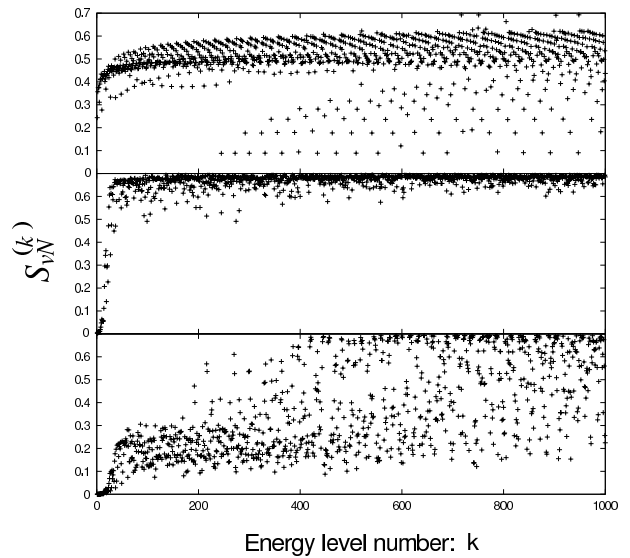}
\caption{
\label{fig:dist_J-dep}
$J$ dependence of entanglement production measured by the von Neumann entropy as 
a function of the energy level number. 
Top: $J=7.5$. Middle: $J=1.5$. Bottom: $J=0.3$. 
The Duschinsky angle is fixed as $\theta=\pi/6$.
}
\end{center}
\end{figure}

\begin{figure}[htbp]
\hfill
\begin{center}
\includegraphics[scale=0.6]{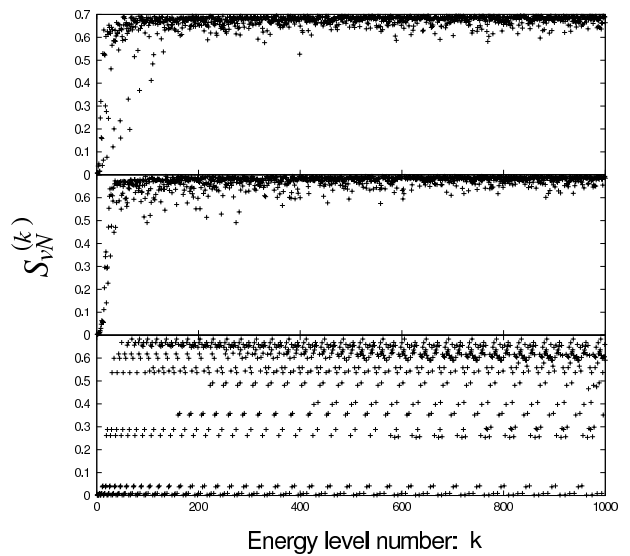}
\caption{
\label{fig:dist_theta-dep}
$\theta$ dependence of entanglement production measured by the von Neumann entropy as 
a function of the energy level number. 
Top: $\theta=\pi/3$. Middle: $\theta=\pi/6$. Bottom: $\theta=0.0$. 
The nonadiabatic coupling is fixed as $J=1.5$.
}
\end{center}
\end{figure}

First we show the $J$ dependence of the results fixing $\theta=\pi/6$:
As we can see in Fig.~{\ref{fig:dist_J-dep}}, 
the entropies for the case of $J=1.5$ assemble around its maximum $S_{\rm vN} \simeq \log 2$,
whereas those of the other cases ($J=0.3, 7.5$) are rather broadly distributed.
This condition of entanglement 
is very similar to that of quantum chaos behavior 
found in \cite{FT01b}: When both $J$ and $\theta$ have ``intermediate'' values 
($J \simeq 1$ and $\theta \simeq \pi/4$), the system shows quantum chaos behavior,
i.e., the nearest neighbor spacing distribution becomes the Wigner type, 
$\Delta_3$ statistics a log curve, and the amplitude distribution 
of the eigenstates Gaussian.
To further confirm this, we show the $\theta$ dependence of 
the results fixing $J=1.5$ in Fig.~{\ref{fig:dist_theta-dep}}. 
This result also nicely corresponds to the previous condition for the 
quantum chaos behavior.
From these results, we can conclude that, 
{\it in a statistical sense}, 
the condition for quantum chaos behavior to arise in the 
TMTS system is very similar to 
that for entanglement production in eigenstates to arise in the same system.
This conclusion result supports the previous study \cite{FNP98} 
which uses entanglement production as an indication of quantum chaos behavior.
(It is also noted that the calculation of entanglement production 
is rather easier than those like the nearest neighbor spacing distibution
or $\Delta_3$ statistics because there is no fitting procedure.)


\begin{figure}[htbp]
\hfill
\begin{center}
\includegraphics[scale=0.6]{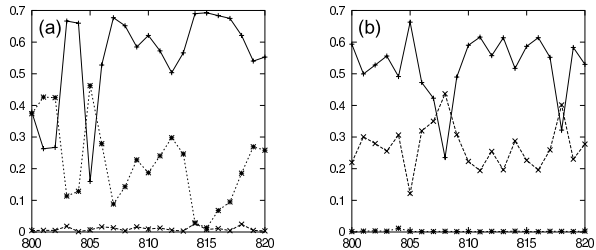}
\caption{
\label{fig:density}
$S_{\rm vN}^{(k)}$ (+), $|\Delta P^{(k)}_{AB}|$ (*), and $|S_{AB}^{(k)}|$ ($\times$) 
as a function of energy level number.
(a) $J=0.3$ (weakly nonadiabatic case).
(b) $J=7.5$ (strongly nonadiabatic case).
(Lines are just for guiding eyes.)
The Duschinsky angle is fixed as $\theta=\pi/6$.
}
\end{center}
\end{figure}

However, as noticed in Fig.~{\ref{fig:dist_J-dep}}, 
the amount of entanglement strongly varies depending on 
{\it each} eigenstate for the cases of $J=0.3$ and 7.5.
To consider this problem, we rewrite the 
entropy (entanglement) using the vibrational bases 
for each electronic state: 
$|\phi_A^{(k)} \rangle$ and $|\phi_B^{(k)} \rangle$.
These states are connected to the above states 
$|\phi_1^{(k)} \rangle$ and $|\phi_2^{(k)} \rangle$ by
\begin{eqnarray}
|\phi_A^{(k)} \rangle
&=& \frac{1}{\sqrt{2}}
(|\phi_1^{(k)} \rangle -|\phi_2^{(k)} \rangle),
\\
|\phi_B^{(k)} \rangle
&=& \frac{1}{\sqrt{2}}
(|\phi_1^{(k)} \rangle +|\phi_2^{(k)} \rangle).
\end{eqnarray}
Hence the density operator, Eq.~(\ref{eq:ele_den}), is 
represented as 
\begin{eqnarray}
\rho^{(k)}_{11}
&=&
\frac{1}{2}+ \langle \phi_A^{(k)}|\phi_B^{(k)} \rangle
\equiv
\frac{1}{2}+ S_{AB}^{(k)},
\\
\rho^{(k)}_{22}
&=&
\frac{1}{2}- \langle \phi_A^{(k)}|\phi_B^{(k)} \rangle
\equiv
\frac{1}{2}- S_{AB}^{(k)},
\\
\rho^{(k)}_{12}
&=&
\frac{1}{2}(\langle \phi_A^{(k)}|\phi_A^{(k)} \rangle
-\langle \phi_B^{(k)}|\phi_B^{(k)} \rangle)
\equiv 
\Delta P^{(k)}_{AB}
\end{eqnarray}
where we have introduced two new parameters: 
$S_{AB}^{(k)}$ is the overlap between 
the $k$-th eigenstates on surface $A$ and $B$,
and $\Delta P^{(k)}_{AB}$ is the half of the population difference 
between the $k$-th eigenstates on surface $A$ and $B$.
Using these parameters, the eigenvalues for the entropy is written as 
\begin{equation}
\lambda_{1,2}^{(k)}=\frac{1}{2} \pm \sqrt{|S^{(k)}_{AB}|^2+ |\Delta P^{(k)}_{AB}|^2}.
\end{equation}
From this relation, 
for the entropy to be large, 
{\it both}  
$|S^{(k)}_{AB}|$ and $|\Delta P^{(k)}_{AB}|$ should be small.
We can numerically confirm this for the strongly ``chaotic'' case ($J=1.5$),
and this property might be derived from the random matrix theory.
For less ``chaotic'' cases ($J=0.3, 7.5$), the situation is different:
As shown in Fig.~\ref{fig:density}, 
there is a strong correlation between $S_{\rm vN}^{(k)}$ and 
$|\Delta P^{(k)}_{AB}|$ for the weakly nonadiabatic case ($J=0.3$),
whereas between $S_{\rm vN}^{(k)}$ and $|S^{(k)}_{AB}|$ 
for the strongly nonadiabatic case ($J=7.5$).
On the other hand,
$|S^{(k)}_{AB}| \simeq 0$ for the former and 
$|\Delta P^{(k)}_{AB}| \simeq 0$ for the latter.
This is interpreted as follows:
For the former, the eigenstates ``reside'' on diabatic surfaces $A$ and $B$ 
which are tilted each other. Thus the overlapping between the eigenstates 
$|S^{(k)}_{AB}|$ becomes small 
because the nodal patterns for the eigenstates are also tilted 
(see Fig.~5 in \cite{FT01b}).  
For the latter, 
the eigenstates ``reside'' on adiabatic surfaces, and the 
amplitudes of them on diabatic surfaces $A$ and $B$ are 
similar (see Fig.~7 in \cite{FT01b}), 
hence $|\Delta P^{(k)}_{AB}| \simeq 0$.

\begin{figure}[htbp]
\hfill
\begin{center}
\includegraphics[scale=0.75]{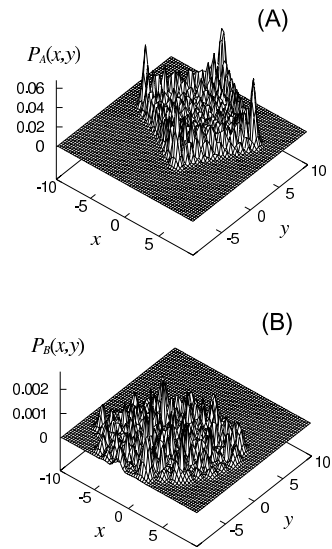}
\caption{
\label{fig:eigen1}
A less entangled regular state:
805-th eigenstates on diabatic surfaces $A$ (A) and $B$ (B) for 
the weakly nonadiabatic case: $J=0.3$.
$P_i(x,y)=|\langle x,y| \phi_i \rangle|^2$ $(i=A,B)$.
Note that the scale for (B) is smaller than that for (A).
The Duschinsky angle is fixed as $\theta=\pi/6$.
}
\end{center}
\end{figure}
\begin{figure}[htbp]
\hfill
\begin{center}
\includegraphics[scale=0.75]{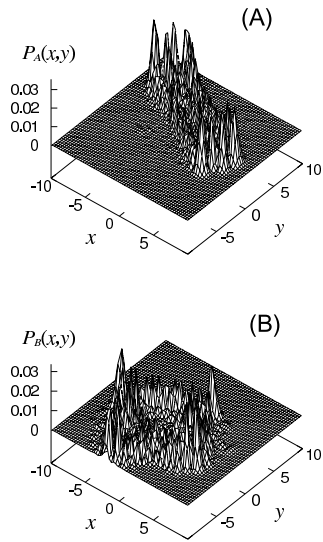}
\caption{
\label{fig:eigen2}
A strongly entangled regular state:
815-th eigenstates on diabatic surfaces $A$ (A) and $B$ (B) 
for the weakly nonadiabatic case: $J=0.3$.
$P_i(x,y)=|\langle x,y| \phi_i \rangle|^2$ $(i=A,B)$.
The Duschinsky angle is fixed as $\theta=\pi/6$.
}
\end{center}
\end{figure}
Let us focus on the weakly nonadiabatic case ($J=0.3$).
In the range of $k=800$ to 820, the lowest entangled 
state is 805-th, and the highest is 815-th
[Fig.~\ref{fig:density} (a)].
In Figs.~\ref{fig:eigen1} and \ref{fig:eigen2},
we show the two eigenstates on diabatic surfaces $A$ and $B$.
As anticipated from the above argument, 
there is a large population difference 
on surface $A$ and $B$ for the less entangled state,
whereas there is not for the strongly entangled state. 
The latter situation means that 
even a regular state can strongly entangle.
Note that, albeit we desymmetrized the system with a 
finite $\Delta \epsilon = \epsilon_A-\epsilon_B$, we 
have this entangled state for the regular case.
(If we do not desymmetrize the system, i.e., $\Delta \epsilon=0$, 
we easily have entangled states for both regular and chaotic cases 
because of the symmetry.)
Thus we must be cautious to use the entanglement production as 
a manifestation of quantum chaotic behavior \cite{FNP98},
though it is fine to use it in a statistical sense.

In this paper, we investigated quantum 
entanglement production between electronic and nuclear (vibrational) 
degrees of freedom for a nonadiabatic system.
We found that the condition of the entanglement in the eigenstates to appear 
is statistically very similar to that of quantum chaos behavior to show up.
We also discussed the nonstatistical behavior of the entanglement production
and interpreted it using the eigenstate properties on diabatic surfaces.
It will be interesting to analyze other nonadiabatic systems like 
Jahn-Teller molecules \cite{KDC84} in light of entanglement production.


\begin{thebibliography}{99}


\bibitem{NC00}
M.A.~Nielsen and I.L.~Chuang,
{\it Quantum Computation and Quantum Information}
(Cambridge University Press, Cambridge, 2000).

\bibitem{PK02}
J.P.~Palao and R.~Kosloff, 
Phys.~Rev.~Lett.~{\bf 89}, 188301 (2002);
Phys.~Rev.~A {\bf 68}, 062308 (2003).

\bibitem{VAZLK02}
J.~Vala, Z.~Amitay, B.~Zhang, S.R.~Leone, and R.~Kosloff,
Phys.~Rev.~A {\bf 66}, 062316 (2002).

\bibitem{Nakamura02}
H. Nakamura, {\it Nonadiabatic Transition: Concepts, Basic Theories and 
Applications} (World Scientific, Singapore, 2002);
C. Zhu, Y. Teranishi, and H. Nakamura, 
Adv.~Chem.~Phys.~{\bf 117}, 127 (2001).


\bibitem{TAWM00}  
K.~Takatsuka, Y.~Arasaki, K.~Wang, and V.~McKoy, 
Faraday Discuss.~{\bf 115}, 1 (2000);
Y.~Arasaki, K.~Takatsuka, K.~Wang, and V.~McKoy,
Phys.~Rev.~Lett.~{\bf 90}, 248303 (2003).


\bibitem{FT01b}
H.~Fujisaki and K.~Takatsuka,
Phys.~Rev.~E {\bf 63}, 066221 (2001). 
See also 
H.~Fujisaki and K.~Takatsuka, 
J.~Chem.~Phys.~{\bf 114}, 3497 (2001);
H.~Higuchi and K.~Takatsuka,
Phys.~Rev.~E {\bf 66}, 035203(R) (2002). 


\bibitem{Heller90}  E.J.~Heller, 
J.~Chem.~Phys.~{\bf 92}, 1718 
(1990).



\bibitem{Gutzwiller90}
M.~C.~Gutzwiller, 
{\it Chaos in Classical and Quantum Mechanics} 
(Springer-Verlag, New York, 1990). 


\bibitem{FNP98}
K.~Furuya, M.C.~Nemes, and G.Q.~Pellegrino,
Phys.~Rev.~Lett.~{\bf 80}, 5524 (1998).

\bibitem{Fujisaki03}
Note, on the other hand, that there is a sound correspondence between 
the TMTS system and its mapping (quasi-classical) system. 
See H.~Fujisaki, quant-ph/0401136; Phys.~Rev.~E (in press).




\bibitem{TLL03}
See 
J.~Tang, M.T.~Lee, and S.H.~Lin,
J.~Chem.~Phys.~{\bf 119}, 7188 (2003),
and references therein.




\bibitem{KDC84}
H.~K{\"o}ppel, W.~Domcke, and L.S.~Cederbaum,
Adv.~Chem.~Phys.~{\bf 57}, 59 
(1984);
D.M.~Leitner, H.~K{\"o}ppel, and L.S.~Cederbaum, 
J.~Chem.~Phys.~{\bf 104}, 434 (1996); 
H.~Yamasaki, Y.~Natsume, A.~Terai, and K.~Nakamura, 
Phys.~Rev.~E {\bf 68}, 046201 (2003);
A.P.~Hines, C.M.~Dawson, R.H.~McKenzie, and G.J.~Milburn,
quant-ph/0402016. 


\end{thebibliography}
\end{document}